\documentstyle{jfm}
\input psfig

\title[Heat transport by turbulent Rayleigh-B\'enard Convection]{Heat transport by turbulent Rayleigh-B\'enard Convection in cylindrical cells with aspect ratio one and less} 
\author{Alexei Nikolaenko, Eric Brown, Denis Funfschilling, and Guenter Ahlers}
\affiliation{Department of Physics and iQUEST,\\ University of
California, Santa Barbara, CA  93106}
\begin{document}

\maketitle

\begin{abstract}
We present high-precision measurements of the Nusselt number $\cal N$ as a function of the Rayleigh number $R$ for cylindrical samples of water (Prandtl number $\sigma = 4.4$) with a diameter $D$ of 49.7  cm and heights $L = 116.3, 74.6$, and 50.6 cm, as well as for $D = 24.8$ cm and $L = 90.2$ cm. For each aspect ratio  $\Gamma \equiv D/L = 0.28, 0.43, 0.67$, and 0.98 the data cover a range of a little over a decade of $R$. The maximum $R \simeq 10^{12}$ and Nusselt number ${\cal N} \simeq 600$ were reached for $\Gamma = 0.43$ and $D = 49.7$. The data were corrected for the influence of the finite conductivity of the top and bottom plates on the heat transport in the fluid to obtain estimates of $\cal N_{\infty}$ for plates with infinite conductivity. The results for ${\cal N}_{\infty}$ and $\Gamma \geq 0.43$ are nearly independent of $\Gamma$. For $\Gamma = 0.275$ ${\cal N}_{\infty}$ falls about 2.5 \% below the other data. For $R \stackrel {<}{_\sim} 10^{11}$, the effective exponent $\gamma_{eff}$ of ${\cal N}_{\infty} = N_0 R^{\gamma_{eff}}$ is about 0.321, larger than those of the Grossmann-Lohse model with its current parameters by about 0.01. For  the largest Rayleigh numbers covered for $\Gamma = $ 0.98, 0.67, and 0.43, $\gamma_{eff}$  saturates at the asymptotic value $\gamma = 1/3$ of the Grossmann-Lohse model. The data do not reveal any crossover to a Kraichnan regime with $\gamma_{eff} > 1/3$.

\end{abstract}

\section{Introduction}
\label{sec:introduction}

Understanding turbulent Rayleigh-B\'enard convection (RBC) in a fluid heated from below [\cite{Si94,Ka01,AGL02}] is one  of the challenging and largely unsolved problems in nonlinear physics.
An important aspect is the global heat transport that is usually expressed in terms of the Nusselt number 
\begin{equation}
{\cal N} = Q L / \lambda \Delta T\ .
\label{eq:nusselt}
\end{equation}
Here $Q$ is the heat-current density, $L$ the sample  height, $\Delta T$ the applied temperature difference, and $\lambda$ the thermal conductivity of the fluid in the absence of convection. A central prediction of various theoretical models [\cite{Si94,GL00,GL01,GL02,GL04}] is a relationship between $\cal N$, the Rayleigh number 
$R = \alpha g \Delta T L^3/\kappa \nu$
 ($\alpha$ is the isobaric thermal expension coefficient, $\kappa$ the thermal diffusivity,  $\nu$ the kinematic viscosity, and $g$ the acceleration of gravity), and the Prandtl number $\sigma = \nu/\kappa$. A model developed recently by \cite{GL00}, based on the decomposition of the 
kinetic and the thermal dissipation into boundary-layer and bulk contributions, provided an excellent fit to experimental data of \cite{XBA00} and \cite{AX01} for a cylindrical cell of aspect ratio $\Gamma \equiv D/L = 1$ ($D$ is the sample diameter) when it was properly adapted [\cite{GL01}, GL] to the relatively small Reynolds numbers of the measurements. However, the data were used to determine five adjustable parameters of the model. Thus more stringent tests using measurements for the same $\Gamma$ but over wider ranges of $R$ and $\sigma$ are desirable. A great success of the model was the excellent agreement with recent results by \cite{XLZ02} for much larger Prandtl numbers than those of \cite{AX01}, at Rayleigh numbers near $1.78\times 10^7$ and $1.78\times 10^9$.  

Here we present new measurements in a cell of diameter $D = 49.7$  cm for $\sigma \simeq 4.4$ that, for $\Gamma = 0.98$,  extend to $R \simeq 10^{11}$. We also report results for $D = 49.7$ cm and $\Gamma = 0.427$ and  0.667, as well as for $D = 24.8$ cm and $\Gamma = 0.275$. For $\Gamma  \stackrel {>}{_\sim} 0.5$ it is expected that the large-scale flow (LSF) in the cell [\cite{KH81}] consists of a single convection roll [\cite{VC03}]. For $\Gamma  \stackrel {<}{_\sim} 0.5$,  on the other hand, \cite{VC03} suggest that the system contains two or more rolls placed vertically one above the other. How this impacts the heat transport was one of the interesting questions to be addressed. Our results suggest that even the cell with $\Gamma = 0.43$ still contained only one roll because the data for $\cal N$ fall on a smooth line drawn through those for the larger $\Gamma$. The $\Gamma = 0.28$ results fall about 2.5\% below all the other data, suggesting a more complicated, perhaps two-roll, structure for the LSF. 

Most of our measurements were made at a mean temperature of 40$^\circ$C, where $\sigma = 4.38$. However, for $\Gamma = 0.67$ we also made measurements at mean temperatures of 50 and 30$^\circ$C, corresponding to $\sigma = 3.62$  and 5.42 respectively. We found a very gentle decrease of $\cal N$ with increasing $\sigma$, approximately in proportion to $\sigma^{-0.044}$.

One of the experimental problems in the measurement of ${\cal N}(R)$ is that the side wall often carries a significant part of the heat current. Corrections for this effect are not easily made, because of the thermal contact between the wall and the fluid that yields a two-dimensional temperature field in the wall, and because of the influence of lateral heat currrents through the wall on the fluid flow [\cite{Ah00,RCCHS01,Ve02,NS03}]. The present project was designed to provide data that are not uncertain due to a significant side-wall correction. We used a classical fluid of relatively large conductivity confined by side walls of relatively low conductivity. The system of choice was water confined by Plexiglas with various heights $L$, and with the greatest diameter permitted by other constraints. We built four convection cells, three with $D = 49.67$ cm with heights $L = 116.33$, 74.42, and 50.61 cm, and one with $D = 24.81$ cm and $L = 90.18$ cm.  For the $\Gamma = 0.98$ ($L = 50.61$) cell, which is most relevant to comparison with the theoretical model of GL, we estimated a wall correction [using Model 2 of \cite{Ah00}] of only 0.3\% for $R = 5\times 10^9 ({\cal N} \simeq 100)$ and smaller corrections for larger $R$. Based on this estimate we felt justified in neglecting the correction.

A second experimental problem was pointed out recently by \cite{CCC02} and by \cite{Ve04}. Using direct numerical simulation, \cite{Ve04} showed that end plates of finite conductivity diminish the heat transport in the fluid when the Nusselt number becomes large. In a separate paper we shall give details of the apparatus used in our work [\cite{NFA04}]. There we will describe measurements using 
two types of top and bottom plates of identical shape and size, but one set made of copper with a conductivity $\lambda_{Cu} = 391$ W/m K and the other of aluminum with $\lambda_{Al} = 161$ W/m K. That work yielded a correction factor that has been applied to the data reported here.

\begin{table}
\begin{center}
\begin{tabular}{cccccccccccccc}
No& $\bar T (^\circ C)$ & $\Delta T (^\circ C)$ & $10^{-8}R$  & $\cal N$ & ${\cal N}_\infty$ & & No& $\bar T (^\circ C)$ & $\Delta T (^\circ C)$ &  $10^{-8}R$  & $\cal N$ & ${\cal N}_\infty$ \\
 1 & 39.932 &  2.122 & 576.8 & 229.1 & 229.5 & &  2 & 39.964 &  4.036 & 1098.4 & 281.4 & 282.3\\
 3 & 40.001 &  5.935 & 1617.2 & 317.2 & 318.6 & &  4 & 40.044 &  7.817 & 2133.1 & 346.6 & 348.6\\
 5 & 40.090 &  9.689 & 2648.5 & 371.1 & 373.6 & &  6 & 40.141 & 11.553 & 3163.5 & 392.4 & 395.4\\
 7 & 40.191 & 13.405 & 3677.1 & 411.4 & 415.0 & &  8 & 39.997 &  2.983 & 812.7 & 255.1 & 255.7\\
 9 & 40.032 &  4.888 & 1333.4 & 298.5 & 299.7 & &  10 & 40.015 &  3.934 & 1072.6 & 278.6 & 279.4\\
 11 & 40.012 &  2.261 & 616.4 & 233.7 & 234.1 & &  12 & 39.990 &  6.448 & 1756.4 & 325.5 & 327.1\\
 13 & 39.994 & 13.406 & 3652.1 & 410.0 & 413.6 & &  14 & 39.973 &  8.941 & 2433.8 & 361.2 & 363.4\\
 15 & 40.020 & 10.809 & 2947.2 & 383.3 & 386.1 & &  16 & 39.998 & 11.833 & 3224.0 & 393.7 & 396.8\\
 17 & 40.017 & 15.710 & 4283.1 & 430.5 & 434.7 & &  18 & 40.007 & 13.773 & 3753.6 & 413.2 & 416.9\\
 19 & 40.031 & 17.636 & 4810.5 & 446.6 & 451.3 & &  20 & 40.002 & 19.647 & 5353.6 & 462.0 & 467.3\\
 21 & 39.997 &  9.872 & 2689.6 & 372.2 & 374.7 & &  22 & 39.997 &  7.908 & 2154.5 & 347.4 & 349.3\\
\end{tabular}
\caption{Results for $\Gamma = 0.275$ and $D = 24.81$ cm.   In this and all following data tables two points are listed per line, and they are numbered in chronological sequence.}
\label{0.28cu}
\end{center}
\end{table}

\begin{table}
\begin{center}
\begin{tabular}{ccccccccccccccc}
No& $\bar T (^\circ C)$ & $\Delta T (^\circ C)$ & $10^{-8}R$  & $\cal N$ & ${\cal N}_\infty$ & &No & $\bar T (^\circ C)$ & $\Delta T (^\circ C)$ &  $10^{-8}R$  & $\cal N$ & ${\cal N}_\infty$ \\
 1 & 39.849 & 19.921 & 11603.4 & 595.4 & 624.7 & &  2 & 39.976 & 17.705 & 10358.7 & 574.5 & 601.6\\
 3 & 40.021 & 15.661 & 9177.2 & 553.1 & 577.9 & &  4 & 40.028 & 13.693 & 8025.7 & 529.2 & 551.7\\
 5 & 39.993 & 11.811 & 6914.1 & 505.3 & 525.4 & &  6 & 39.996 &  9.843 & 5763.1 & 476.9 & 494.6\\
 7 & 40.092 &  7.689 & 4517.2 & 441.1 & 455.8 & &  8 & 40.008 &  3.936 & 2305.5 & 356.1 & 364.8\\
 9 & 40.009 &  1.965 & 1151.2 & 286.0 & 291.0 & &  10 & 40.053 &  2.862 & 1679.0 & 322.4 & 329.2\\
 11 & 40.057 &  4.824 & 2830.3 & 380.6 & 390.8 & &  12 & 39.979 &  5.958 & 3486.2 & 405.8 & 417.8\\
 13 & 40.014 &  1.464 & 857.9 & 260.2 & 264.1 & &
  \end{tabular}
\caption{Results for $\Gamma = 0.427$ and $D = 49.7$ cm.}
\label{0.43cu}
\end{center}
\end{table}

\begin{table}
\begin{center}
\begin{tabular}{ccccccccccccccc}
No& $\bar T (^\circ C)$ & $\Delta T (^\circ C)$ & $10^{-8}R$  & $\cal N$ & ${\cal N}_\infty$ & &No& $\bar T (^\circ C)$ & $\Delta T (^\circ C)$ &  $10^{-8}R$  & $\cal N$ & ${\cal N}_\infty$ \\
 1 & 39.922 & 17.639 & 2693.9 & 371.5 & 389.3 & &  2 & 39.816 & 16.035 & 2439.8 & 359.5 & 376.0\\
 3 & 39.923 & 13.881 & 2120.1 & 343.9 & 358.8 & &  4 & 39.937 & 11.892 & 1817.2 & 327.7 & 341.0\\
 5 & 39.984 &  7.891 & 1207.7 & 288.0 & 297.9 & &  6 & 39.995 &  3.955 & 605.5 & 231.5 & 237.3\\
 7 & 40.013 &  1.952 & 299.1 & 185.8 & 189.1 & &  8 & 39.979 & 19.633 & 3004.5 & 384.2 & 403.4\\
 9 & 39.964 & 20.637 & 3156.4 & 389.9 & 409.7 & &  10 & 39.898 & 18.821 & 2872.0 & 378.4 & 396.9\\
 11 & 39.984 & 16.693 & 2555.0 & 365.1 & 382.2 & &  12 & 39.915 & 14.875 & 2271.3 & 351.4 & 367.1\\
 13 & 39.890 & 12.974 & 1979.2 & 336.5 & 350.6 & &  14 & 40.109 & 10.574 & 1625.6 & 316.0 & 328.3\\
 15 & 39.974 &  8.900 & 1361.7 & 299.4 & 310.2 & &  16 & 40.059 &  2.846 & 436.8 & 209.0 & 213.5\\
 17 & 39.974 & 17.693 & 2707.1 & 371.7 & 389.5 & &  18 & 49.989 & 19.566 & 4101.7 & 426.0 & 450.1\\
 19 & 49.886 & 18.789 & 3927.2 & 420.2 & 443.6 & &  20 & 49.969 & 17.651 & 3698.0 & 412.3 & 434.8\\
 21 & 49.959 &  5.003 & 1047.9 & 277.7 & 286.8 & &  22 & 50.087 &  6.710 & 1410.6 & 304.8 & 316.0\\
 23 & 50.136 &  8.576 & 1805.4 & 329.9 & 343.4 & &  24 & 50.146 & 10.503 & 2211.6 & 351.2 & 366.8\\
 25 & 50.174 & 12.406 & 2614.4 & 370.4 & 388.1 & &  26 & 50.097 & 14.525 & 3054.1 & 388.4 & 408.1\\
 27 & 49.902 &  4.140 & 865.7 & 261.6 & 269.4 & &  28 & 49.935 &  3.092 & 647.1 & 239.0 & 245.3\\
 29 & 49.928 &  2.120 & 443.6 & 212.4 & 217.0 & &  30 & 49.973 &  1.046 & 219.1 & 172.3 & 175.0\\
 31 & 39.974 & 17.693 & 2707.1 & 371.7 & 389.5 & &  32 & 29.980 & 19.647 & 2007.0 & 336.1 & 350.3\\
 33 & 29.996 & 17.656 & 1805.0 & 325.1 & 338.2 & &  34 & 29.841 & 16.006 & 1624.5 & 314.4 & 326.6\\
 35 & 29.742 & 14.246 & 1439.0 & 302.2 & 313.2 & &  36 & 29.913 & 11.963 & 1218.2 & 287.0 & 296.8\\
 37 & 29.957 &  9.918 & 1012.1 & 270.5 & 278.9 & &  38 & 29.932 &  8.011 & 816.6 & 252.8 & 260.0\\
 39 & 29.859 &  6.190 & 628.8 & 232.5 & 238.3 & &  40 & 29.897 &  4.157 & 423.0 & 205.0 & 209.2\\
 41 & 29.904 &  2.171 & 221.0 & 167.0 & 169.5 & &  42 & 30.036 &  2.898 & 296.8 & 183.6 & 186.8\\
 43 & 30.011 &  4.913 & 502.6 & 216.6 & 221.5 & &  44 & 30.018 &  6.858 & 701.8 & 240.6 & 246.9\\
 45 & 29.992 &  2.491 & 254.6 & 174.3 & 177.1 & & & & & & & \\
  \end{tabular}
\caption{Results for $\Gamma = 0.667$ and $D = 49.7$ cm.}
\label{table:0.67cu}
\end{center}
\end{table}

\begin{table}
\begin{center}
\begin{tabular}{ccccccccccccccc}
No& $\bar T (^\circ C)$ & $\Delta T (^\circ C)$ & $10^{-8}R$  & $\cal N$ & ${\cal N}_\infty$ & & No& $\bar T (^\circ C)$ & $\Delta T (^\circ C)$ &  $10^{-8}R$  & $\cal N$ & ${\cal N}_\infty$ \\
 1 & 40.092 &  1.792 & 86.7 & 125.3 & 127.5 & &  2 & 40.006 &  2.940 & 141.7 & 145.6 & 148.9\\
 3 & 40.015 &  3.898 & 188.0 & 158.8 & 162.9 & &  4 & 39.975 &  4.951 & 238.4 & 170.9 & 175.7\\
 5 & 39.953 &  5.969 & 287.2 & 181.2 & 186.8 & &  6 & 39.933 &  6.979 & 335.6 & 190.2 & 196.4\\
 7 & 39.933 &  7.952 & 382.3 & 197.9 & 204.8 & &  8 & 39.966 &  9.832 & 473.3 & 211.4 & 219.4\\
 9 & 39.935 & 11.829 & 568.8 & 224.2 & 233.4 & &  10 & 39.916 & 13.803 & 663.3 & 235.1 & 245.4\\
 11 & 40.039 & 15.500 & 748.1 & 244.3 & 255.5 & &  12 & 39.407 & 16.728 & 789.6 & 248.3 & 259.9\\
 13 & 39.933 & 17.645 & 848.4 & 254.1 & 266.3 & &  14 & 40.038 &  2.386 & 115.1 & 136.2 & 138.9\\
 15 & 39.996 &  1.489 & 71.8 & 118.0 & 119.9 & &  16 & 39.935 & 17.638 & 848.2 & 254.0 & 266.3\\
 17 & 39.943 & 17.633 & 848.2 & 254.2 & 266.4 & & & & & & & \\  
 \end{tabular}
\caption{Results for $\Gamma = 0.981$ and $D = 49.7$ cm.}
\label{table:1.0cu}
\end{center}
\end{table}

\section{Results}
\label{sec:results}

\subsection{The measurements}

Details of the apparatus and of experimental procedures are given by \cite{NFA04}. The precision of the measurements is typically near 0.1 percent, and systematic errors, primarily due to uncertainties of the diameter and length of the cell, are estimated to be well below one percent. Deviations from the Boussinesq approximation are believed to be unimportant  [\cite{NFA04}]. We give the results in Tables~\ref{0.28cu} to \ref{table:1.0cu}. The measured  $\cal N$ derived from Eq.~\ref{eq:nusselt} as well as the Nusselt number ${\cal N}_\infty$ obtained after correction for the finite top- and bottom-plate conductivity [\cite{Ve04}] are listed. The relation between these is given by
${\cal N} = f(X) {\cal N}_\infty$
where $X$ is the ratio of the average thermal resistance of the plates to that of the fluid.  We used the empirical function
$f(X) = 1 - exp[-(a X)^{b}]$,
with the parameters $a = 0.275$ and $b = 0.390$ for $D = 49.7$, and $a = 0.304$ and $b = 0.506$ for $D = 24.82$, determined experimentally [\cite{NFA04}]. The parameters were found to be independent of the aspect ratio but to depend on the plate diameter.

\begin{figure}
\centerline{\psfig{file=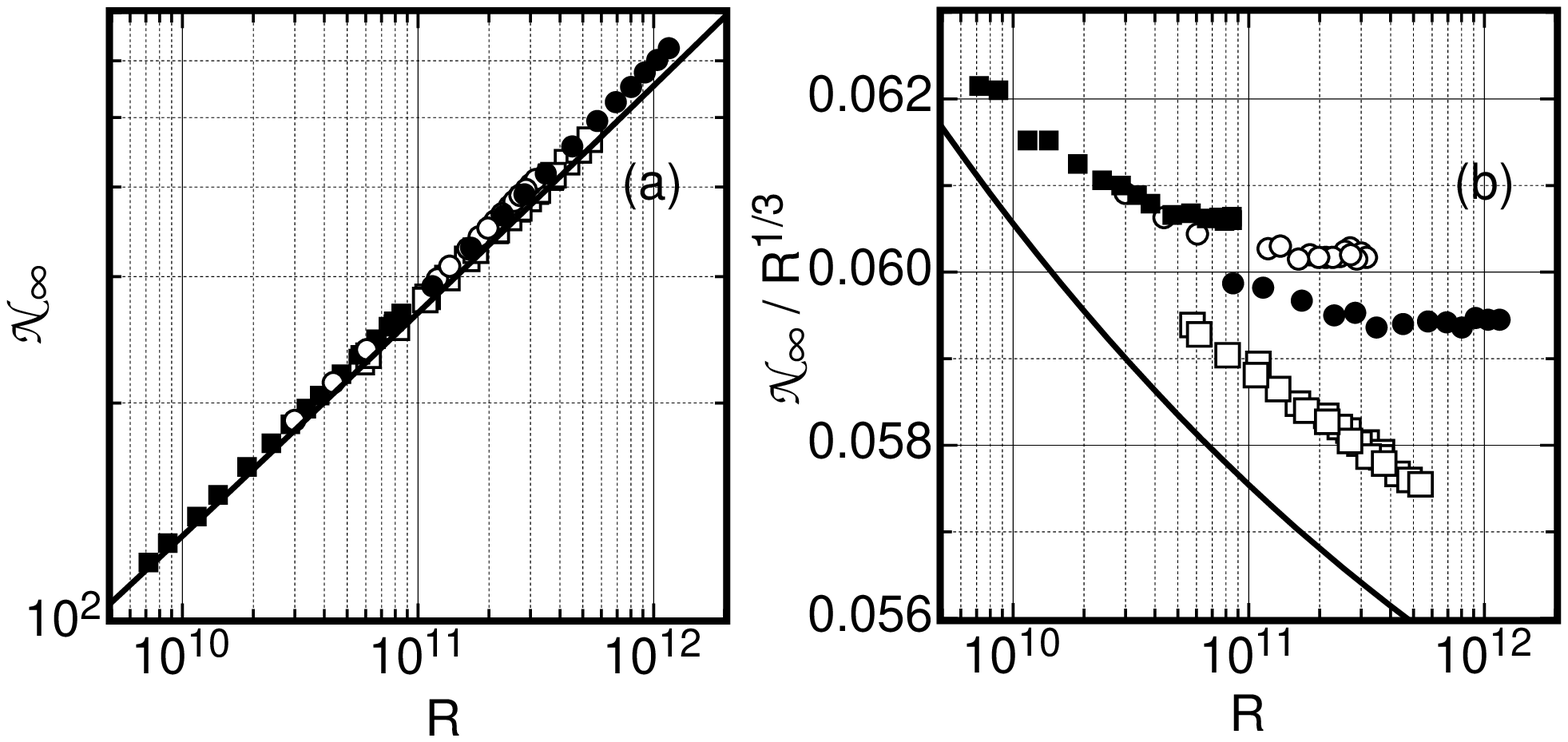,height=2.5in}}
\caption{(a): The Nusselt number ${\cal N}_\infty$ as a function of the Rayeigh number $R$ on logarithmic scales. (b): The compensated Nusselt number ${\cal N}_\infty /R^{1/3}$ as a function of the Rayleigh number $R$.  Open squares: $\Gamma = 0.275$. Solid circles: $\Gamma = 0.43$. Open circles: $\Gamma = 0.67$. Solid squares: $\Gamma = 0.98$. Solid lines: the prediction of \protect \cite{GL01} for $\sigma = 4.4$.}
\label{fig:nusselt}
\end{figure}

\subsection{Dependence on $R$ and $\Gamma$}

The results for ${\cal N}_\infty$ are shown on logarithmic scales in Fig.~\ref{fig:nusselt}a and with greater resolution in the compensated form ${\cal N_\infty}/R^{1/3}$  in Fig.~\ref{fig:nusselt}b. We note that over most of the range of $R$ the data for $\Gamma \geq 0.43$  reveal very little if any dependence of ${\cal N}_\infty$ on $\Gamma$. Recent considerations by \cite{GL03} had suggested a stronger $\Gamma$-dependence. Earlier experimental data [\cite{WL92,XBA00}] also suggested a stronger $\Gamma$ dependence; but those results were influenced by side-wall and/or end-plate effects. Note for instance that the side-wall correction made by \cite{Ah00} considerably reduced the $\Gamma$ dependence originally seen by \cite{XBA00}. Similarly, our present results for $\cal N$ reveal some dependence on $\Gamma$, but the end-plate correction largely removes it. It is particularly noteworthy that the data for $\Gamma = 0.43$ are not shifted significantly relative to those for $\Gamma = 0.67$ because on the basis of the numerical calculations of \cite{VC03} one expects different structures for the LSF for these two cases. The data for $\Gamma = 0.275$ are lower by about 2.5\%, suggesting that a transition in the LSF structure may occur between $\Gamma = 0.43$ and 0.275.

\begin{figure}
\centerline{\psfig{file=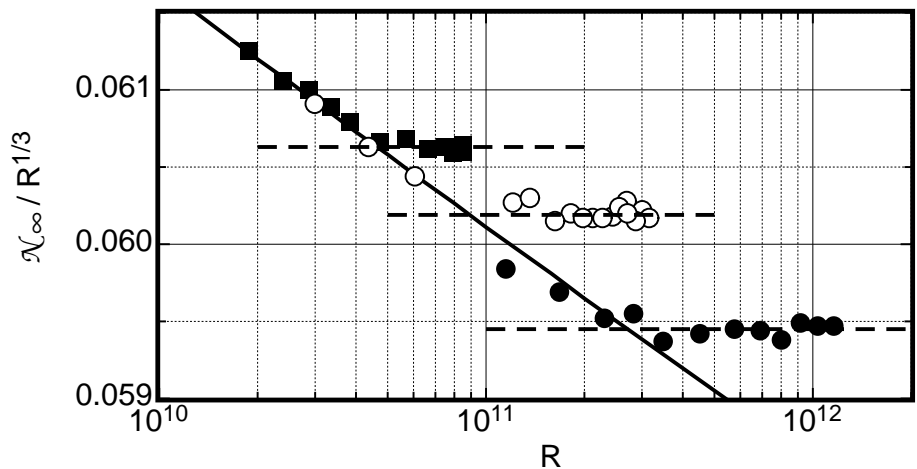,height=2.5in}}
\caption{The compensated Nusselt numbers ${\cal N}_\infty/R^{1/3}$ as a function of the Rayleigh number $R$ for $\Gamma = 0.98$ (solid squares), $\Gamma = 0.67$ (open circles), and $\Gamma = 0.43$ (solid circles) on an expanded scale. Lines: powerlaws ${\cal N}_\infty = N_0 R^{\gamma_{eff}}$. Solid line: $N_0 = 0.0797, \gamma_{eff} = 0.3222$. Dashed lines: $\gamma_{eff} = 1/3$ and (from top to bottom) $N_0 = 0.06063, 0.06019, 0.05945$.}
\label{fig:nusselt2}
\end{figure}

A second important feature of the data is their dependence on $R$. Locally, over a limited range, the measurements can be fit by the powerlaw
\begin{equation}
{\cal N}_\infty = N_0 R ^ {\gamma_{eff}}\ .
\end{equation}
For the effective exponent we found $\gamma_{eff} = 0.323$ for $\Gamma = 1$ near $R = 2\times10^{10}$ and $\gamma_{eff} = 0.329$ for $\Gamma = 0.67$ near $R = 10^{11}$. As shown in Fig.~\ref{fig:nusselt2}, a single fit to most of the data for $\Gamma = 0.98, 0.67$, and 0.43 yields $\gamma_{eff} = 0.322$. All these values are close to, but definitely less than, the asymptotic large-$R$ prediction $\gamma = 1/3$ of the GL model. However, they are larger by about 0.01 or 0.015 than the GL prediction for $\Gamma = 1$ at the same $R$. This can also be seen qualitatively from Fig.~\ref{fig:nusselt}b where the GL prediction with its present parameter values is shown as a solid line. It remains to be seen whether the model parameters can be adjusted so as to reproduce this feature.

Another interesting aspect noticeable in Fig.~\ref{fig:nusselt}b is that for  $\Gamma = 0.98$, 0.67, and 0.43 there is a seemingly sudden change in the $R$-dependence of ${\cal N}_\infty$ at large $R$ to a power law with $\gamma = 1/3$, i.e. with the asymptotic prediction of the GL model. This is illustrated more clearly in Fig.~\ref{fig:nusselt2} which shows the relevant data with  higher resolution.  It is difficult to see how the GL model can reproduce this rather sudden transition  at finite $R$ to its asymptotic exponent value. Rather, it seems likely that a new physical phenomenon not yet contained in the model will have to be invoked. The transition occurs for $\Gamma = 0.98, 0.67$, and 0.43 near $R = 5\times 10^{10}, 9\times 10^{10}$, and $3\times 10^{11}$ respectively and is reflected in the observation that the data for larger $R$ fall on horizontal lines in the figures. In this range the measurements do reveal a dependence on $\Gamma$, with $N_0(\gamma_{eff}=1/3) = 0.0606, 0.0602$, and 0.0595 for $\Gamma = 0.98, 0.67$, and 0.43 respectively. One interpretation of these results for $N_0$ is  that the heat transport is diminished, albeit only very slightly, by a larger travel distance, and presumably a larger period, of the LSF.  At smaller $R$ an effective powerlaw with $N_0 = 0.0797, \gamma_{eff} = 0.322$ (solid line in Fig.~\ref{fig:nusselt2}) fits the data at all three $\Gamma$ values within their statistical uncertainty, showing that the results are essentially $\Gamma$ independent. It is a surprise that the data for $\Gamma = 0.67$ and 0.43 do not differ more from each other because different structures for the LSF had been predicted for these two cases. [\cite{VC03}] The $\Gamma = 0.275$ results do indeed have a Nusselt number that is smaller by about 2 to 3\%, suggesting that the transition in the flow structure occurred between $\Gamma = 0.43$ and 0.275. Those result do not show the saturation at large $R$ with $\gamma_{eff} = 1/3$, and lead to $\gamma_{eff} = 0.321$ over their entire range.

\begin{figure}
\centerline{\psfig{file=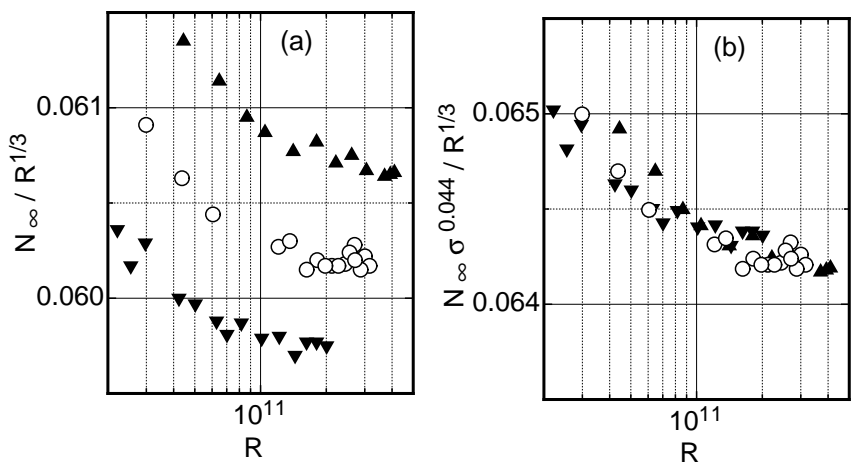,height=2.5in}}
\caption{(a): The compensated Nusselt numbers ${\cal N}_\infty/R^{1/3}$ as a function of the Rayleigh number $R$ for $\Gamma = 0.667$ at three values of the Prandtl number. (b): The reduced Nusselt numbers ${\cal N}_\infty \sigma^{0.044}/R^{1/3}$ as a function of $R$. Up pointing triangles: 50$^\circ$C and $\sigma = 3.62$. Open circles: 40$^\circ$C and $\sigma = 4.38$. Down-pointing  triangles: 30$^\circ$C and $\sigma = 5.42$.}
\label{fig:N_of_prandtl}
\end{figure}

\subsection{Dependence on the Prandtl number $\sigma$}

The Nusselt number ${\cal N_\infty}(\sigma)$ has a broad maximum near $\sigma\simeq 4$. Thus the dependence of $\cal N$ on $\sigma$ is very weak and difficult to determine from measurements with various fluids of different $\sigma$
because of systematic errors due the uncertainties of the fluid properties [\cite{AX01,XLZ02}]. We determined ${\cal N}(\sigma,R)$ with high precision over a narrow range of  $\sigma$ by using 
the same convection  cell  and by changing the mean temperature. In that case errors from different cell geometries are largely absent and the properties are known very well. Similar measurements over the ranges 
$5\times 10^6 \leq R \leq 5\times 10^8$ and $4 \leq \sigma \leq 6.5$ were made by \cite{LE97}. Their results can be represented well by 
\begin{equation}
{\cal N} = N_{00} \sigma^{\alpha_{eff}} R ^ {\gamma_{eff}}
\label{eq:powerlaw}
\end{equation}
with $\gamma_{eff} = 0.286$, $\alpha_{eff} = -0.030$, and $N_{00} = 0.1780$. The negative value of $\alpha_{eff}$ indicates that the maximum of ${\cal N}(\sigma)$ is below $\sigma \simeq 4$.

We used our $\Gamma = 0.67$ cell and made measurements at 30, 40, and 50$^\circ$C corresponding to $\sigma =  5.42$, 4.38, and 3.62 respectively. The results are included in Table ~\ref{table:0.67cu}, and in the compensated form ${\cal N}_\infty/R^{1/3}$ they are shown as a function of $R$ in Fig.~\ref{fig:N_of_prandtl}a. Over the range $2\times 10^{10} \leq R \leq 4\times 10 ^{11}$ they yield $\alpha_{eff} \simeq -0.044$, indicating that also for this range of $R$ the maximum of ${\cal N}(\sigma)$ occurs below $\sigma = 4$. The reduced Nusselt number ${\cal N}_\infty \sigma^{0.044}/R^{1/3}$ is shown in Fig.~\ref{fig:N_of_prandtl}b. It shows that the data, within their experimental uncertainty, collapse onto a unique curve when divided by $\sigma^{\alpha_{eff}}$. The observed $\sigma$-dependence is somewhat stronger than that of the GL model with its present parameter values.

\begin{figure}
\centerline{\psfig{file=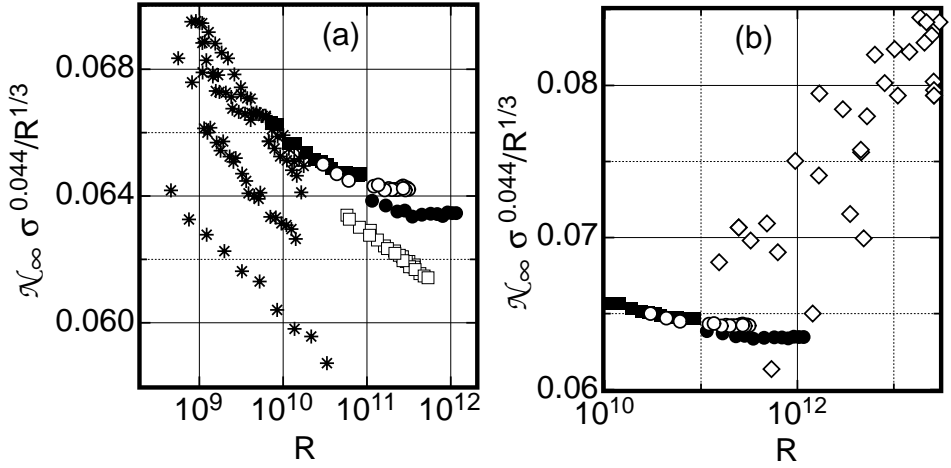,height=2.5in}}
\caption{The reduced Nusselt numbers ${\cal N}_\infty \sigma^{0.044}/R^{1/3}$ as a function of $R$. (a): Stars: Data of \cite{RCCH04} for $2.5 \leq \sigma \leq 6.0$ and $\Gamma = 0.50$. Other symbols: this work for $\sigma = 4.4$ and $\Gamma = 0.98$ (solid squares), 0.67 (open circles), 0.43 (solid circles), and 0.28 (open squares). (b): open diamonds: data of \cite{CCCCH01} for $\Gamma = 0.5$ and $2.0 \leq \sigma \leq 4.5$; solid symbols and open circles: this work for $\Gamma = 0.98$, 0.67, and 0.43.}
\label{fig:others}
\end{figure}

\subsection{Comparison with other results}

Measurements of ${\cal N}(R)$ for $\Gamma = 0.50$ over the range $10^8 \leq R \leq 5\times 10^{10}$ and a wide range of $\sigma$ were made at cryogenic temperatures using gaseous helium by \cite{RCCH04}. A direct, highly quantitative comparison with our results is possible only for the data points with $\sigma$-values fairly close to ours where the $\sigma$-dependence of $\cal N$ can reasonably be expected to be given by Eq.~\ref{eq:powerlaw} with $\alpha_{eff} \simeq -0.044$. In  Fig.~\ref{fig:others}a we show results of \cite{RCCH04} for ${\cal N}_\infty \sigma^{0.044}/R^{1/3}$ as a function of $R$. These data were corrected by the authors for the side-wall conductance, using a procedure described by them. Because of the small conductivity of helium gas one expects end-plate corrections to be negligible in this case.  We also show our results for comparison. One sees that the data of \cite{RCCH04} fall into three well defined groups, with a nearly uniform vertical spacing between them of close to four percent.  There is excellent agreement/consistency of the uppermost branch with our data for $\Gamma = 0.98$, 0.67, and 0.43. This is consistent with the absence of any significant aspect-ratio dependence in our data. The existence of the lower two branches is more difficult to reconcile with our results. \cite{RCCH04} suggest that they encountered more than one distinct state of their LSF. In our work we never found multi-stability for any $\Gamma$, and the data for $\Gamma = 0.67$ and 0.43 (which according to the calculations of \cite{VC03} should correspond to different flow structures) agree with each other and are consistent with the upper branch of Roche et al., at least in the range where $\gamma_{eff}$ has not yet saturated at $\gamma_{eff} = 1/3$. Our data for $\Gamma = 0.28$ are lower than those for our larger $\Gamma$, suggesting a difference in the LSF and indicating that the transition from a single cell to a more complicated structure occurs at $\Gamma < 0.43$ in our system. However, our results for $\Gamma = 0.28$ are only about  2 or 3 \% lower than the larger-$\Gamma$ data and not as low as the results from the middle or lower branch of Roche et al. 

In Fig.~\ref{fig:others}b we compare our results with those of \cite{CCCCH01} from cryogenic experiments for $2.0 \leq \sigma \leq 4.5$ and $\Gamma = 0.50$.  The end-plate corrections are expected to be negligible. Because of the low conductivity of the fluid, the side-wall contribution to the conductance of the cell is significant, but apparently no correction was made. The authors interpret their data to imply $\gamma_{eff} \simeq 0.38$, and attribute the large value to a breakdown of the boundary layers adjacent to the top and bottom plates. In that case one expects that a new regime first proposed by \cite{Kr62} with an asymptotic exponent $\gamma = 1/2$ (and logarithmic corrections) should be entered. Our data do not reveal such a large exponent, and in the overlapping range $10^{11} \stackrel {<}{_\sim} R \stackrel {<}{_\sim} 2\times 10^{12}$ remain consistent with $\gamma_{eff} = 1/3$. 
 
\begin{figure}
\centerline{\psfig{file=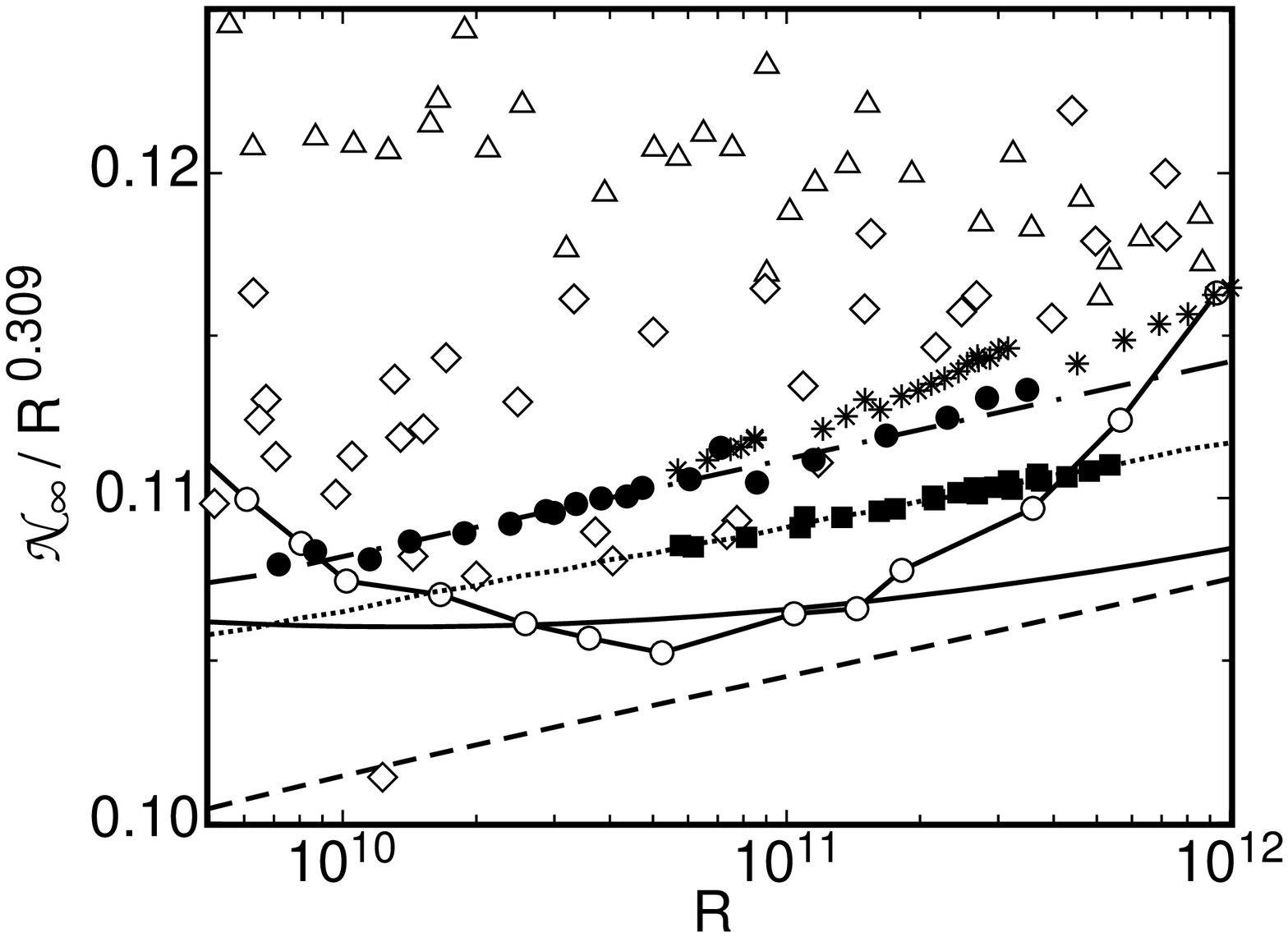,height=2.5in}}
\caption{Compensated Nusselt numbers ${\cal N}_\infty /R^{0.309}$ as a function of $R$.
Open circles: \protect \cite{NS03}, $\Gamma = 1, 0.7 < \sigma < 0.9$. Open triangles: \protect \cite{NSSD00},  $\Gamma = 1/2, \sigma \simeq 0.7$. Open diamonds: \protect \cite{CCCCH01},  $\Gamma = 1/2, 0.7 < \sigma < 2.0$. Solid circles and stars: present work, $0.43 \leq \Gamma \leq 0.98, \sigma = 4.4$. Solid squares: present work, ${\cal N}_\infty, \Gamma = 0.28, \sigma = 4.4$. Dash-dotted line: powerlaw fit to our data with $\Gamma \geq 0.43$ ($\gamma_{eff}= 0.3207$). Dotted line: powerlaw fit to our data for $\Gamma = 0.28$ ($\gamma_{eff}= 0.3193$). Solid (dashed) line: GL model for $\sigma = 4.4 ~ (0.8)$.}
\label{fig:heliumdata}
\end{figure}

Finally, in Fig.~\ref{fig:heliumdata} we compare our results for $\Gamma = 0.98, 0.67,$ and 0.43 (solid circles  where $\gamma_{eff} \simeq 0.32$ and stars where $\gamma_{eff} \simeq 0.333$) and $\Gamma = 0.28$ (solid squares) on a less sensitive vertical scale with several measurements at cryogenic temperatures. A more comprehensive comparison was presented by \cite{NS03} (NS) [we choose the same representation in terms of ${\cal N}_\infty /R^{0.309}$ that was used by them in their Fig.~5]. There are systematic differences between the data sets that are, according to  NS, larger than possible systematic errors in the experiments. Our data show very little change of $\gamma_{eff}$ with $R$, and the solid circles in Fig.~\ref{fig:heliumdata} yield $\gamma_{eff} \simeq 0.321$. 
 The data by NS (connected open circles) show a dependence on $R$ that differs from ours, with $\gamma_{eff}$ varying from about 0.28 to about 0.35 as $R$ changes from $10^9$ to $10^{12}$.  The data of \cite{NSSD00} do not show such a strong variation of $\gamma_{eff}$ and can be described well by a single $\gamma_{eff} = 0.30$ over the $R$-range of the figure. The data of \cite{CCCCH01} (open diamonds) are on average slightly larger than ours but have an $R$-dependence that is consistent with that of ours.
 
We have been unable to rationalize the  large variations of the effective exponents of $\cal N$ from one experiment to another, and in the case of the data of NS within  a single experiment with $R$, in terms of  known fluid mechanics. NS suggested that differences in the LSF structure are responsible, and state that ``such differences seem to arise from delicate interplay among detailed geometry as well as Prandtl and Rayleigh numbers". Looking at our data, the absence of any  significant $\Gamma$-dependence is apparent from the clustering of the data for $\Gamma \geq 0.43$ (solid circles) near the straight dash-dotted line corresponding to $\gamma_{eff} = 0.3207$; i.e. $\Gamma$ {\it per se} does not play a major role. Even for $\Gamma = 0.28$, where the data are displaced vertically in the figure, presumably because of a change in the LSF, they have nearly the same effective exponent $\gamma_{eff} = 0.3193$ (dotted line). Thus it appears from our  data that differences in the LSF do not have a large effect  on the $R$-dependence of ${\cal N}_\infty(R)$. We wish we had a more satisfying explanation of the different results obtained by the various experiments.

\section{Acknowledgment}

This work was supported by the US Department of Energy through Grant  DE-FG03-87ER13738.

\end{document}